\documentclass[12pt]{revtex4}
\usepackage[]{graphicx}
\usepackage{amsmath}
\usepackage{amsfonts}
\usepackage[brazil]{babel} 
\usepackage[latin1]{inputenc} 
\usepackage{tikz}
\usetikzlibrary{decorations.text}
\usetikzlibrary{shapes.geometric}
\usetikzlibrary{arrows}
\usepackage{amssymb}
\usepackage{makeidx}
\usepackage{epstopdf}
\usepackage{subfigure}

\newcommand{\beq}{\begin{equation}} \newcommand{\eeq}{\end{equation}}
\newcommand{\bea}{\begin{eqnarray}} \newcommand{\eea}{\end{eqnarray}}
\newcommand{\bear}{\begin{eqnarray*}} \newcommand{\eear}{\end{eqnarray*}}

\begin{document}

\title {Mean Field Cluster approximation scheme for the duplet-creation model with absorbing phase transition}

\author{A. A. Ferreira}

\affiliation{Departamento de  F\'\i sica, Universidade Federal de S\~ao Paulo, 96010-900  S\~ao Paulo, S\~ao Paulo, Brazil}
\email{ferreira07@unifesp.com.br}

%
%

\begin{abstract}
  We study the nonequilibrium phase transitions in the one-dimensional duplet creation model using the $n-$site approximation scheme.  We find  the phase diagram in the space of parameters $(\gamma,D)$, where $\gamma$  is the particle decay probability and $D$ is the diffusion probability. Through data $(1\leq n \leq 18)$ we
  show that in the limit $n\rightarrow \infty$  the model presents a continuous transition of active state for inactive state (absorbing state) for any value of $0\leq D \leq1$.  In general, we obtain the  critical value of $\gamma$ and the ``gap'' density in the transition point for  single and pair approximation $\gamma_c(n=1,D)$, $\Delta\rho(n=1,D)$ and $\Delta \rho(n=2,D)$ respectively.
\end{abstract}


\maketitle

\section{Introduction}
\label{sec:intro}

In recent decades there has been a great interest of Statistical Mechanics about one-dimensional systems that exhibit a continuous ~\cite{MK16A1,MK16A10} or discountinous ~\cite{MK16A11,MK16A15} phase transitions far from equilibrium. In particular there are a class of models that presents a discontinuous phase transition from an active to an inactive state ~\cite{MK16A16,MK16A18}.

Although the same kind of transition has beeen observed in duplet creation model ~\cite{MK17}, this result, however, is not accepted by the statistical physics
community in view of a general argument due  Hinrichsen ~\cite{MK18} that first-order transitions cannot occur in fluctuating one-dimensional systems because the surface tension of a domain
does not depend on its size. 

In this paper we review tha phase diagram of the one-dimesional  duplet creation model with diffusion  ~\cite{MK17}, trought the $n-$site approximation  method (mean field cluster analysis) \cite{MK19}. We shall show that this model exhibit a continuous phase transition from an active phase to a unique asborving phase (the vacum state) belonging to the (DP) universality classes.
 
 This result is interesting because, as was observed in the triplet model  \cite{MK20}, the $n-$site approximation scheme was not conclusive to confirm the absence of a tricritical point ~\cite{MK21}.
In this way, it is natural to expect that the duplet model with diffusion also presents a continuous phase transition.

 The rest of the paper is organized as follows. In section 2 we describe the duplet creation model, whereas in section 3 the single-site ($n=1$)
and the pair site approximation ($n=2)$ are discussed analytically, however the general case ($n\geq3$) is just treated numerically. Finally, in section 4 we discuss the results and the limitations of the $n-$site approximation scheme.

\section{The Model}
\label{sec:1}

This model can be seen as a  interacting hard core particles models lattice with three processes: spontaneous annihilation (with rate $\gamma$), creation particles by 
two particles (with rate $s$), and diffusion of particles (with rate $D$). The parameters $D$, $s$ and $\gamma$ 
are such that $D+\gamma +s=1$. 

The configuration of sites in lattice  is represented by the vector $\left|\beta^n\right\rangle\equiv\left|\beta_1\right\rangle\otimes\left|\beta_2\right\rangle\otimes\ldots\otimes\left|\beta_n\right\rangle$, where $n$ is the number os sites of lattice.
 We assume that $\beta_{i}$ takes on the value 1 if the site $i$ is 
occupied by a particle and the value $0$ if the site is empty. The evolution rules of model are as follows.

A site, for example, $i$ is chosen randomly among the $n$ sites of the lattice. Suppose $\beta_{i}=1$,
then there are two possible actions:  the particle decays with probability $\gamma$, so  the site $i$ becomes 
vacant, or it moves to a $2$ neighboring site with probability $D/2$. Of course, the site $i$ remains unchanged 
with probability $s=1-\gamma -D$.

Next, suppose that the site $i$ is empty, i.e.,  $\beta_{i}=0$. The first step is to choose with equal probability $(1/2)$ one of the 
directions (right or left). Suppose the right side is chosen, as previously, there are two possibilities: occupation of
site $i$ with probability $s$, provided that $\beta_{i+1} = \beta_{i+2} = 1$. If the two sites which are neighbors of site $i$,  are occupied, then with probability $s$ a new particle is created,  or  with probability $D$ the variables $\beta_i$ and $\beta_{i+1}$ are interchanged. A
similar procedure is applied in the case the neighborhood at the left side of $i$ is chosen.

\section{The $n$-site approximation}

Writing the master equation in its continuous-time differential form, we have

\begin{eqnarray}
\label{eq:master1}
\partial t \left|P({\sigma},t)\right\rangle=\sum_{\beta}w_{(\beta\rightarrow\sigma)}\left|P({\beta},t)\right\rangle-
w_{(\sigma\rightarrow\beta)}\left|P({\sigma},t)\right\rangle ,
\end{eqnarray}

  \noindent{where $\sigma,\beta$ represent two distinct lattice configuration. 
Rewriting  the equation ($\ref{eq:master1}$) in its vector form \cite{MK36}, }

\begin{eqnarray}
\label{eq:master2}
\partial_t \left|P\right\rangle=-H\left|P\right\rangle
\end{eqnarray}

\noindent{where $H$ is a matrix operator, responsible for connecting differents configurations of the vector space. 
It is also important to mention that, in general, this operator is not Hermitian, i.e., it has complex eigenvalues. 
These eingenvalues correspond to the oscillations in the model (imaginary part), while the 
exponential decay is contained in the real part.}

\noindent{In an orthonormal basis we have $\left\langle \sigma^n\right|\left|\beta^n\right\rangle=\delta_{\sigma_1,\beta_1}\delta_{\sigma_2,\beta_2}\cdots\delta_{\sigma_n,\beta_n}$. This suggests that we can
write $|P>$ as}

\begin{eqnarray}
\left|P\right\rangle=\sum_{\beta}P(\beta,t)\left|\beta\right\rangle.
\end{eqnarray}

  \noindent{If we denote the initial probability of the system by $\left|P_o\right\rangle=\sum_{\beta}P_o(\beta)\left|\beta\right\rangle,$ 
the formal solution of the problem can be written as }

\begin{eqnarray}
\left|P\right\rangle=e^{-Ht}\left|P_o\right\rangle.
\end{eqnarray}

  Due to conservation of probability, we have $\left\langle 0\right|H=0$, where $\left\langle 0\right|=\sum_{\beta}\left\langle \beta\right|$. Thus any
observable can be calculated as follows

\begin{eqnarray}
&&<X>_t=\sum_{\beta}X(\beta)P(\beta,t)\left|\beta\right\rangle=\nonumber\\
&&\left\langle 0\left| X\right|P \right\rangle=\left\langle 0\left| Xe^{-Ht}\right|P_o \right\rangle.
\end{eqnarray}

However, to compute this amount is necessary to diagonalize the evolution operator $H$. This task
  it is not always viable, since the dimension of this operator $H$  grows like $b^n$, where $b$ is the number of states of a site $i$.

To circumvent this difficulty, some numerical procedures are usually adopted, such as Monte Carlo simulations,
numerical diagonalization of the operator $H$ through DMRG scheme, pertubative expansion and others techniques ~\cite {MK18}. Here, we make approximations in the components of the vector $\left|P\right\rangle$ \cite{MK19} and compute the time evolution of it.

 We present now, just as we did in \cite{MK20}, a special scheme  to obtain the 
discrete time evolution of the Master Equation. Since the process is Markovian and the site update rule are independent of $t$ we can write the component
of equation (\ref{eq:master1}) as

\begin{eqnarray}
\label{eq:f1}
 P_n(\left|\sigma^n\right\rangle;t+\Delta t)=\sum_{\left|\beta^n\right\rangle}W_{\Delta t}(\left|\beta^n\right\rangle\rightarrow\left|\sigma^n\right\rangle)P_n(\left|\beta^n\right\rangle;t).
\end{eqnarray}

\noindent{Here $W_{\Delta t}(\left|\beta^n\right\rangle\rightarrow\left|\sigma^n\right\rangle)$ is the conditional probability
of the transition from configuration $\left|\beta^n\right\rangle$ to configuration $\left|\sigma^n\right\rangle$ in the time interval $\Delta t$. We choose
this discrete-time formulation rather than the usual continuous-time approach in order to preserve the interpretation of the parameters
 $\gamma , D$ and $s$ as probabilities. Using $W_{\Delta t}(\left|\beta^n\right\rangle\rightarrow\left|\sigma^n\right\rangle)=
1-\sum_{\left|\beta^n\right\rangle\neq\left|\sigma^n\right\rangle}$, we can rewrite $(\ref{eq:f1})$ in a more conveniente form 

\begin{eqnarray}
\label{eq:f2}
&& \delta P_n(\left|\sigma^n\right\rangle;t)=\nonumber\\
&&\sum_{\left|\beta^n\right\rangle}[W_{\Delta t}(\left|\beta^n\right\rangle\rightarrow\left|\sigma^n\right\rangle)P_n(\left|\beta^n\right\rangle;t)-
W_{\Delta t}(\left|\sigma^n\right\rangle\rightarrow\left|\beta^n\right\rangle)P_n(\left|\sigma^n\right\rangle;t)],\nonumber\\
\end{eqnarray}

\noindent{where $\delta P_n(\left|\sigma^n\right\rangle;t)=P_n(\left|\sigma^n\right\rangle;t+\Delta t)-P_n(\left|\sigma^n\right\rangle;t)$. As usual, the continuous-time 
formulation is obtained by dividing both sides of equation $(\ref{eq:f2})$ by $\Delta t$, taking the limit $\Delta t\rightarrow 0$, and
defining $W_{\Delta t}(\left|\beta^n\right\rangle\rightarrow\left|\sigma^n\right\rangle)/\Delta t=w(\left|\beta^n\right\rangle\rightarrow\left|\sigma^n\right\rangle)\equiv w_{(\beta\rightarrow\sigma)}$ as the transition rate between configurations $\left|\beta^n\right\rangle$ and 
$\left|\sigma^n\right\rangle$.}

For finite chain sizes, application of the site update rules $W_{\Delta t}(\left|\beta^n\right\rangle\rightarrow\left|\sigma^n\right\rangle)$ with periodic boundary conditions (i.e., setting $\sigma_{n+1}=\sigma_1$ and $\sigma_0=\sigma_n$), allows that  the dynamics visits 
any configurations beginning from an abritrary initial configuration distinct from  the absorbing steady-state $\left|\sigma^n\right\rangle=\left|0\right\rangle$ (i.e.,the configuration for which $\sigma_i=0$ for 
$i=1,\ldots,n).$ So the unique steady-stade solution of equation ($\ref{eq:f1}$) is $P_n(\left|\sigma^n\right\rangle=\left|0\right\rangle;t\rightarrow \infty)=1.$ In the limit of infinitely large chains $n\rightarrow \infty$,
 a second stable stationary solution of equation ($\ref{eq:f1}$) appears, the so-called active state, for which the average density of
particles $\rho$ is nonzero.

 The basic point now is to describe the stochastic dynamics of a 
$n-$site spin configuration only in terms of the joint probability distribution $P_n(\left|\sigma^n\right\rangle;t)$ using translationary invariant equations. The condition of translational
invariant requires that the update rules for the sites close to the boundaries of the chain are the same as for the inner sites. To achieve that we need to introduce ``virtual'' \cite{MK20}
 \rm  sites, say $i=-1,0$ if the neighborhood of site $i=1$ is considered. The $n$-site approximation is a prescription
to write the $m$-joint probability distributions $(m>n)$ in terms of $P_n(\left|\sigma^n\right\rangle;t)$ only. The basic assuption involved in this approximation scheme is that the states of any two
sites are considered as statistically independent variables if their distance is larger than $n$. For example, we can write the $n+2$ jointed distribution $P_{n+2}(\left|\sigma_{-1},\sigma_{0},\sigma^{n-2}\right\rangle;t)$ as

\begin{eqnarray}
\label {eq:ap_1}
Y_ {n+2}=&&\frac{P_n(\left|\sigma_{-1},\sigma_{0},\sigma^{n-2}\right\rangle)}{P_{n-1}(\left|\sigma_{0},\sigma^{n-2}\right\rangle)}\times\frac{P_n(\left|\sigma_{0},\sigma^{n-1}\right\rangle)}{P_{n-1}(\left|\sigma^{n-1}\right\rangle)}\times P_n(\left|\sigma^{n}\right\rangle),
\end{eqnarray}

\noindent{where the $n-1$-site distribution can be easily written in terms of the $n$-site distribution (we have omitted the dependence on $t$ to lighten the notation).

\begin{eqnarray}
&& P_{n-1}(\left|\sigma_{0},\sigma^{n-2}\right\rangle)=\sum_{\sigma_{-1}=0}^{1}P_n(\left|\sigma_{-1},\sigma_{0},\sigma^{n-2}\right\rangle),\;\;\;\;\;\mbox{and}\\
 &&P_{n-1}(\left|\sigma^{n-1}\right\rangle)=\sum_{\sigma_{0}=0}^{1}P_n(\left|\sigma_{0},\sigma^{n-2}\right\rangle).
\end{eqnarray}

\noindent{Recalling the update rules duplet model: when the site $i=1$ is empty (i.e.,$\sigma_1=0$) 
and  its left neighborhood is chosen for the occupation procedure, then it is necessary that its $2$ virtual neighbors  $i=-1,0$ are occupied (i.e, $\sigma_{-1}=\sigma_{0}=1)$.
In addition, we note that expression $(\ref{eq:ap_1})$ is valid for $n > 2$ only.}

Now we consider the task of updating the vacant site $i=2$ (i.e.,$\sigma_{2}=0)$ in a configuration $|\sigma^n>$ where sites $i=1$ is ocupied. We need to consider $1$ virtual sites $i=0$	 and the relevant
joint distribution $Y_{n+1}\equiv P_{n+1}(\left|\sigma_{0},\sigma^{n}\right\rangle)	$ is given by		

\begin{eqnarray}
\label {eq:ap_2}
Y_ {n+1}=\frac{P_n(\left|\sigma_{0},\sigma^{n-1}\right\rangle)}{P_{n-1}(\left|\sigma^{n-1}\right\rangle)}\times P_n(\left|\sigma^{n}\right\rangle),
\end{eqnarray}

In what follows we present the explicit form of the equations that determine the joint distribution $P_n(\left|\sigma^{n}\right\rangle)$ for $n=1$ and $n=2$, referred to as single-site and pair-approximation respectively. In
both cases we derive analytical expressions for the transition point lines and for the jump in the particle density at the transition. For $n\geq 3$ we computer the numerical solution of equation
$(\ref{eq:f2})$ for the steady-state condition $\delta P_{n}(\left|\sigma^{n}\right\rangle)=0$. We solve those $2^n$ coupled equations using Newton-Raphson method with the requisite of an error smaller than $10^{-16}$ per equation.

\subsection{The single-site approximation}

The relevant quantity is $P_1(1)=\rho$, and $P_1(0)=1-\rho$ is given by normalization condition. Recalling that the only ``real'' site is $i=1$ and we introduce
the convention to write the states of the ``virtual'' sites $(i.e.,i=-1,0,1,2,3)$ with an overlying bar. Therefore we can rewrite equation 
$(\ref{eq:f2})$ in the form

\begin{eqnarray}
 \delta P_1(1)=-\gamma P_1(1)+\frac{s}{2}P_{3}(\bar{1},\bar{1},0)+\frac{s}{2}P_{3}(0,\bar{1},\bar{1}).
\end{eqnarray}

\noindent{The diffusion parameter $D$ does not appear explicity in this equation because its contribution comes from terms such as $DP_2(\bar{1},0)$ and
$-DP_2(\bar{0},1)$ which cancel out because of the parity symmetry. Since in this case the sites are statistically independent we can write 
$P_{3}((\bar1),(\bar1),0)=P_1^{2}(\bar1)P_1(0)$ and similarly for the contribution of the right neighborhood of site $i=1$, so that the last 
equation can be write}

\begin{eqnarray}
\label{eq:eqK}
\delta \rho=-\gamma \rho+ s(1-\rho)\rho^{2}.
\end{eqnarray}

\noindent{The nontrivial solution at stationary state are given by the roots of equation $(1-\rho)\rho=\frac{\gamma}{s}.$ There is one discontinuous point of particle density $\Delta \rho$ at the transition between the active and absorbing phases.
We find $\Delta \rho=\frac{1}{2}$ regardless of the values of the control parameters $\gamma$ and $D$. Inserting the value $\rho=\frac{1}{2}$
in the last equation with $\delta \rho=0$ we find the critical value for $\gamma$.}

\begin{eqnarray}
\label{eq:gcri}
\gamma _c=\frac{(1-D)}{5}. 
\end{eqnarray}

\subsection{The pair approximation}

In the case $n=2$, equation $(\ref{eq:f2})$ can be reduced to only two independent equations using  the parity symmetry
$P_2(0,1;t)=P_2(1,0;t)$ and their normalization conditions. To 
ilustrate the reasoning that leads to the equation ($\ref{eq:ap_1}$) and ($\ref{eq:ap_2}$), we will derive the equation for $P_2(1,1)$
explicity. The $(\ref{eq:f2})$ yields 

\begin{eqnarray}
\label{eq:pp2}
 \delta P_2(1,1)=&&-\gamma P_2(1,1)-\frac{D}{4}P_3(1,1,\bar{0})-\frac{D}{4}P_3(\bar{0},1,1)\nonumber\\
&&+\frac{D}{4}P_3(\bar{1},0,\bar{1})+\frac{D}{4}P_3(1,0,\bar{1})+\frac{s}{4}P_{4}(\bar{1},\bar{1},0,1)\nonumber\\
&&+\frac{s}{4}P_{4}(1,0,\bar{1},\bar{1})+\frac{s}{4}P_{3}(0,1,\bar{1})+\frac{s}{4}P_{3}(\bar{1},1,0),
\end{eqnarray}

\noindent{where, as before, we use the convention of writing the virtual site states with an overlying bar. The factor 1/4 appears here because
the probability of choosing a given site for update is 1/2 (there are only two real sites) and the probability that the left (or the right) 
neighborhood of that site is selected to verify the possibility of diffusion (site interchange) or creation is also 1/2.} 

We begin by working out with expression expression $P_{4}(\bar{1},\bar{1},0,1), P_{4}(1,0,\bar{1},\bar{1}),P_{3}(1,1,\bar{0})$ and  $P_{3}(\bar{1},0,\bar{1})$. We can rewrite theses probabilities distribution as}

\begin{eqnarray}
&&P_{4}(\bar{1},\bar{1},0,1)= \frac{P_2(\bar{1},\bar{1})}{P_1(\bar{1})}\frac{P_2(\bar{1},0)}{P_1(0)}P_2(0,1),\\
&&P_{4}(\bar{1},\bar{1},1,0)= \frac{P_2(\bar{1},\bar{1})}{P_1(\bar{1})}\frac{P_2(\bar{1},1)}{P_1(1)}P_2(1,0),
\end{eqnarray}

\begin{eqnarray}
&&P_{3}(1,1,\bar{0})=P_{3}(\bar{0},1,1)=P_{3}(\bar{1},1,0)=P_{3}(0,1,\bar{1})=\frac{P_2(\bar{0},1)}{P_1(1)}P_2(1,1),\\
&&P_{3}(\bar{1},0,\bar{1})=P_{3}(1,0,\bar{1})=\frac{P_2(\bar{1},0)}{P_1(0)}P_2(0,\bar{1}).
\end{eqnarray}

At this point, we use the parity symmetry to write $\delta P_2(1,1)$ given by equation	($\ref{eq:pp2}$) in terms of $P_2(1,1)$
and $P_2(1,0)$ only. It is still necessary to derive an equation for $\delta P_2(1,0)$, but this can be done quite straightorwardly using the procedure described above.
The final dynamic equations for the pair approximation, posed in terms of usual variables $\varphi=P_2(1,1)$ and $\rho=P_1(1)=P_2(1,1)+P_2(1,0)=\varphi+P_2(1,0)$, are

\begin{eqnarray}
\label{eq:re}
2\delta \rho=-\gamma\rho +s(\rho-\varphi)\frac{\varphi}{\rho},
\end{eqnarray}

\begin{eqnarray}
\label{eq:re2}
2\delta \varphi=-2\gamma\varphi +s\frac{\varphi}{\rho}(\rho-\varphi)(1+\frac{\rho-\varphi}{1-\rho})-
D(\varphi-\rho^2)\frac{\rho-\varphi}{\rho(1-\rho)}.
\end{eqnarray}

Differently of single-site approximation, here the diffusion parameter $D$ introduce a nontrivial contribution to component of the master equation.
The stationary regime is now obtained from solution of ($\ref{eq:re}$) and ($\ref{eq:re2}$) which determines the point of the phase transition. In fact, the reduced variable
$\phi=\varphi/\rho$ is given by the same equation discussed in the single-site approximation and so $\phi=\frac{1}{2}$ at the transition 
line. This implies that the equation of the transition line $\gamma_c(D)$ is also identical to the obtained in the single-site approximation.
However, the size of the discontinuity $\Delta \rho$ at the transition differs in the two approximation schemes. Imposing the steady-steady
condition in last equation yields

\begin{eqnarray}
 \Delta \rho=\frac{1}{4}\frac{D}{(\gamma_c+D)},
\end{eqnarray}

\noindent{where $\gamma_c$ is given by $(\ref{eq:gcri})$. We note that the equations for the transitions line $\gamma_c(D)$
coincides only in the cases of the single-site and pair approximation.  The gap density $\Delta \rho$ vanishes in the pair approximation when $D=0$.

\subsection{The general n-site aproximation}

When $n\geq3$ we have to resort a numerical implementation of equation $(\ref{eq:f2})$. In particular, the configurations
$|\sigma^n>$ (i.e.,the arguments of the joint distribution $P_n$) are represented by $n$ bit integers, which allows an easy
implementation of the boundary sites update rules by the Fortran 95 bit intrinsic functions. We choose an initial configuration such that $P(|\sigma^n=1>;0)\approx 1$ in order to bia the Newton-Raphson method to find the steady-state
 solution of equation  $(\ref{eq:f2})$. 

In the absence of diffusion ($D=0$), the  the one-site approximation fail to predict the continuous phase transition between
 the absorbing and active phases. However, when $n\geq13$, the extrapolation data for $\gamma_c^{\infty}$ converge to $0.12011\pm 0.00001$ whereas the Monte Carlo simulation predict $\gamma_c^{MC}=0.11971$.

\begin{figure}[htb]
	\centering
	\includegraphics[width=10cm]{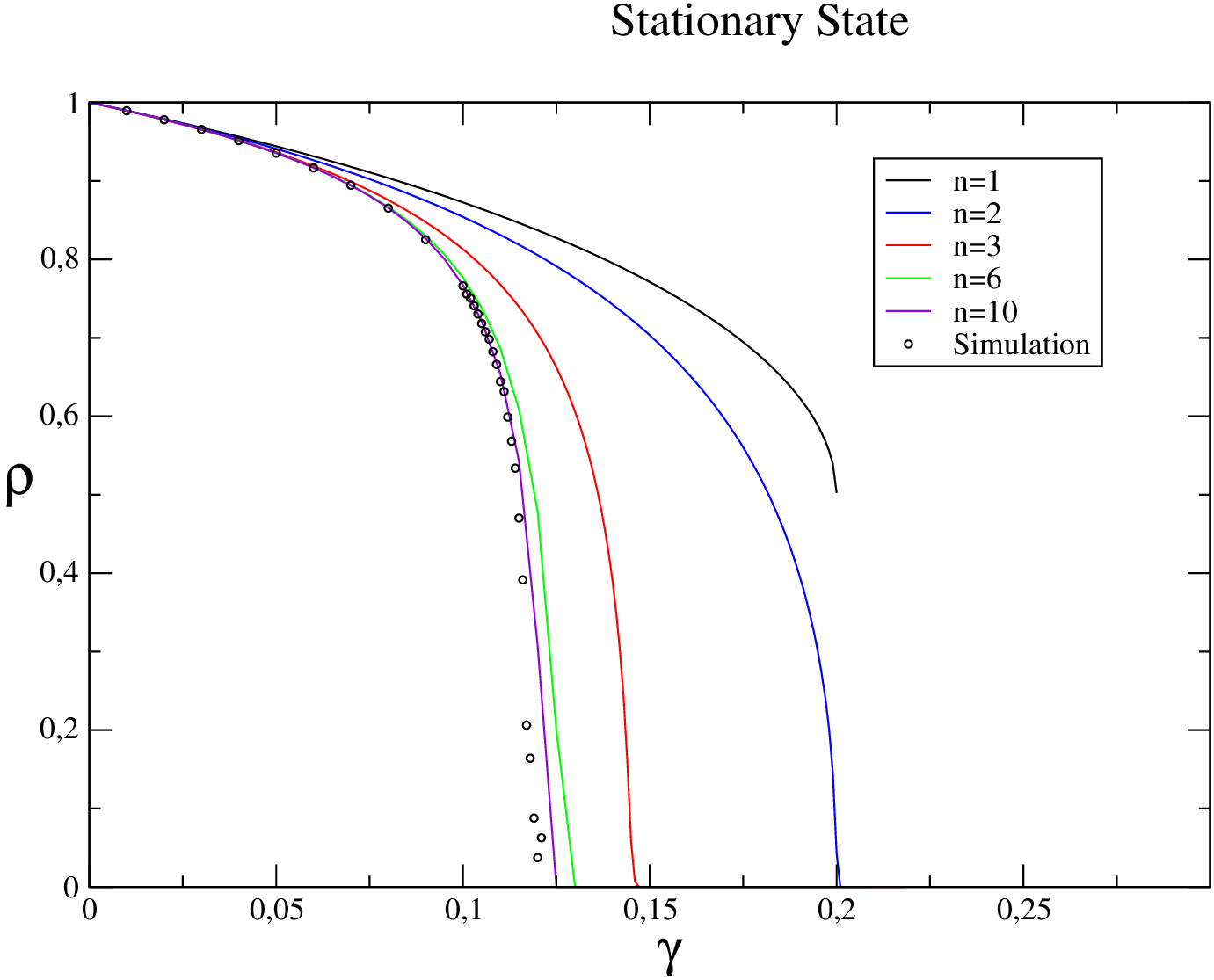}
      \caption{ (color on line) Phase diagram for duplet model with $D=0$. The density of particles at equilibrium $\rho$ as a fuction of the decay probability $\gamma$ and (right to left) $n=1,2,3,6,10$. The symbol $\circ$ are the results of the Monte Carlo simulation.}	
	\label{fig:fig8}
\end{figure}

\begin{figure}[htb]
	\centering
	\includegraphics[width=10cm]{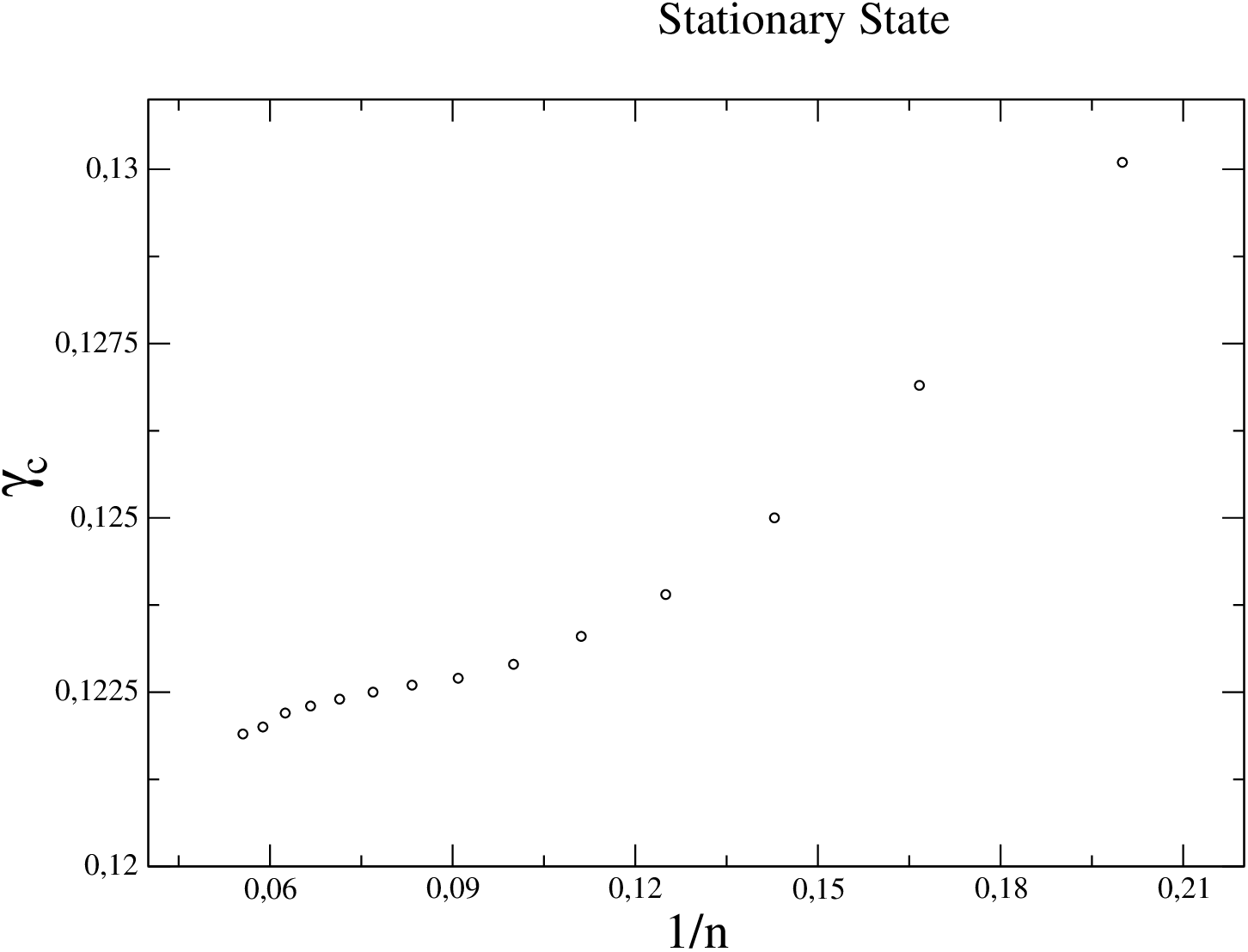}
      \caption{Dependence of the critical value of the decay probability $\gamma$ at which the density of particles vanishes continuously on the order n of the n-site approximation for duplet with $D=0$.}
    \label{fig:fig9}
\end{figure}

In the presence of diffusion ($D\neq0$) , the $n-$site 
approximation (for finite $n$) apparently shows the existence of a tricritical point for different values of $D$. However in the limit $(n\rightarrow \infty)$ the scenario is somewhat more complicated. Because when we 
plot the estimates of the tricitrical point coordinates as a function of the order $1/n$ of $n-$site approximation (figure $\ref {fig:fig11}$ and figure $\ref {fig:fig12}$) we obatin,
by extrapolation of data, the nonphysical estimate for $\gamma_t^{\infty}=-0.03865$ and $D_t^{\infty}=1.34941$. Thus, we can conclude that the duplet model
displays only a continuous phase transition between the absorbing and active phases regardless of the value of the diffusion probability $D$ in disagreement with \cite{MK17}.

\begin{figure}[htb]
      \centering
	\includegraphics[width=10cm]{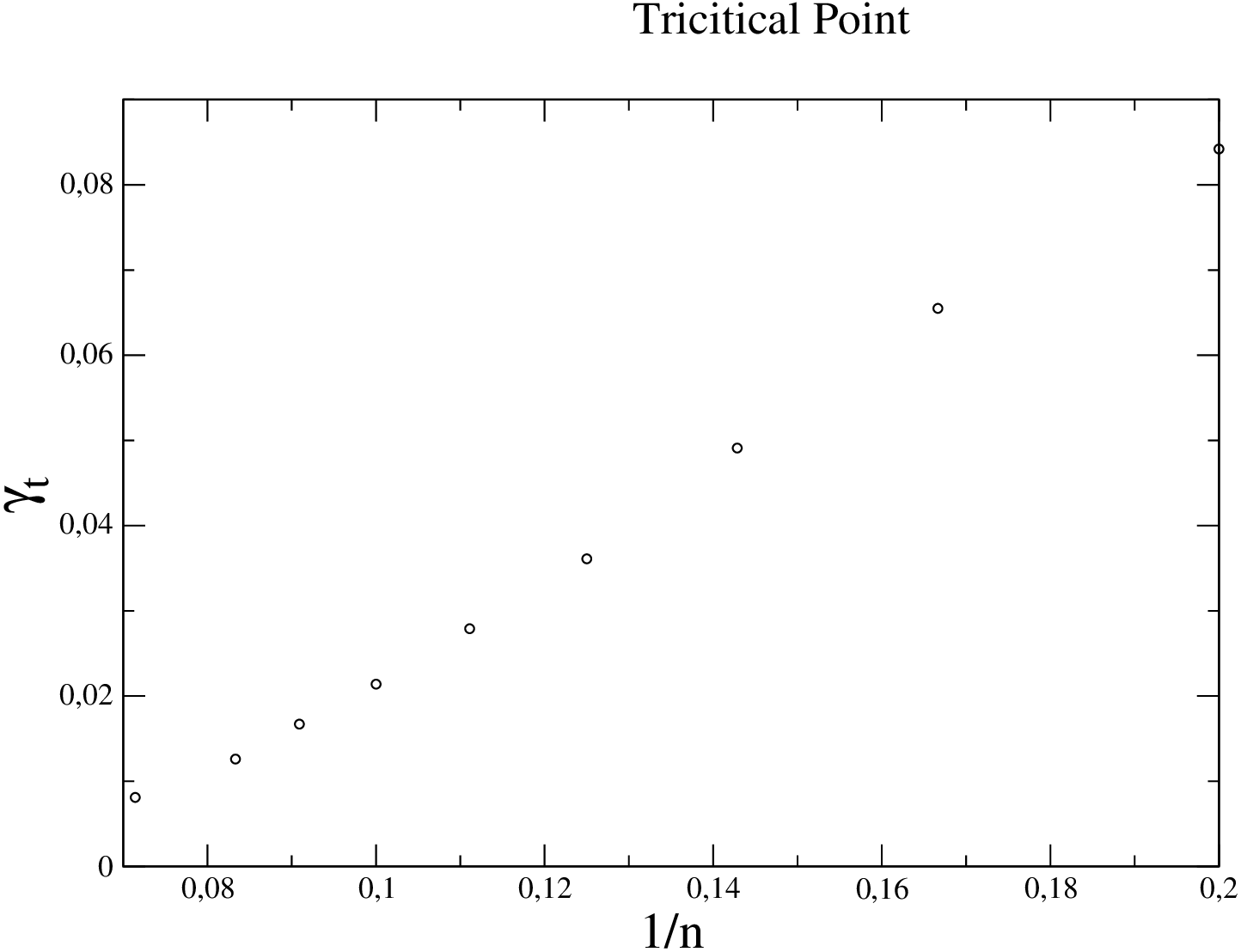}
\caption{Dependence of the value of the tricritical decay probability at which the jump $\delta\rho$ of the density of particles vanishes continuously on the order n of the n-site approximation.}
\label{fig:fig11}
\end{figure}

\begin{figure}[htb]
	\centering
	\includegraphics[width=10cm]{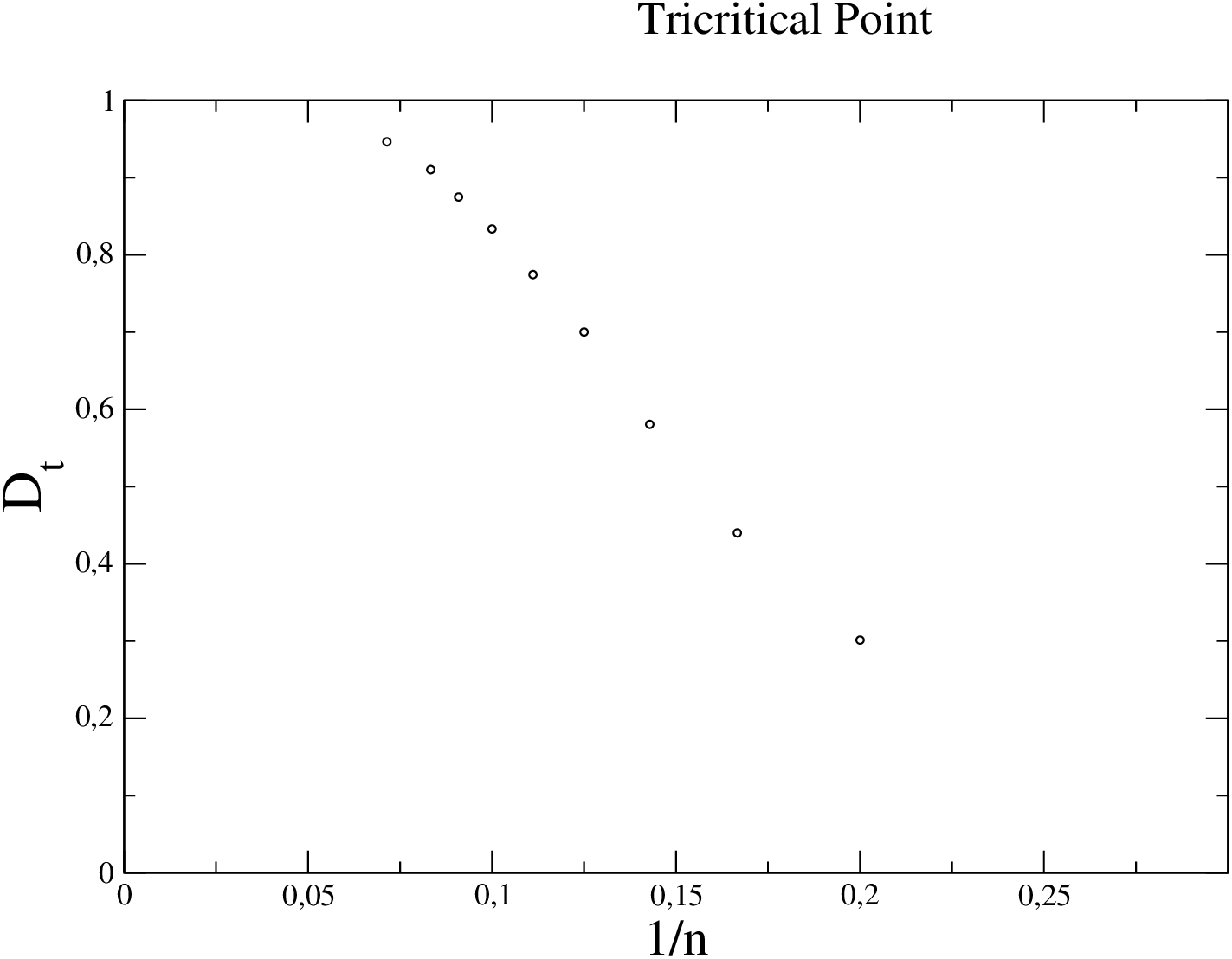}
\caption{Dependence of the value of the tricritical diffusion probability at which the jump $\delta\rho$ of the density of particles vanishes continuously on the order $n$ of the n-site approximation.}
\label{fig:fig12}
\end{figure}

\section{Conclusion}
\label{sec:conclusion}

We investigated the one-dimensional duplet creation model with diffusion \cite{MK17} through the $n-$site approximation scheme \cite{MK19}. We found that in the absence of diffusion $(D=0)$ the $\gamma_c^{\infty}(D=0)=0.12011$, which agrees very well with Monte
Carlo Simulation $\gamma_c^{MC}(D=0)=0.11971$. However, when $D\neq 0$ we observed that the tricritical point is localized in non physical regime
$\gamma_t=-0.03865$ and $D_t=1.3494$ suggesting that the model does not have a tricritical point.

In the general, we obtain for single-site and pair approximation the exactly expression for $\gamma_c(n=1,D)$ and $\Delta_c(n=2,D)$ respectively. Overall we observed that the $n-$site approximation scheme is limited to determine the discontinous phase transitions for systems with long range interactions.

\section{Acknowledgments}

We thank J. F. Fontanari for fruitfull discussions. This work  has been partially supported  by the Brazilian  agencie FAPERGS.

\end{document}